\documentclass[%
 reprint,
 amsmath,amssymb,
 aps,
pre,
]{revtex4-2}

\usepackage{graphicx}% Include figure files
\usepackage[caption=false]{subfig}
\usepackage{dcolumn}% Align table columns on decimal point
\usepackage{bm}% bold math
\usepackage{color}
\usepackage{booktabs}
\usepackage{float}
\usepackage{algpseudocode}
\usepackage{caption}
\usepackage{etoolbox}
\usepackage{acro}

\DeclareAcronym{ML}{short=ML, long=Machine Learning}
\DeclareAcronym{ANNs}{short=ANNs, long=Artificial Neural Networks}
\DeclareAcronym{FC}{short=FC4, long=4-layers fully connected artificial neural network}
\DeclareAcronym{CNN}{short=CNN5, long=5-layers convolutional neural network}
\DeclareAcronym{EM}{short=EM, long=Evolution Metric}
\DeclareAcronym{OE}{short=OE, long=Outcome Entropy}
\DeclareAcronym{PE}{short=PE, long=Prediction Entropy}
\DeclareAcronym{ACC}{short=A, long=Accuracy}

\newcounter{algorithm}

\DeclareCaptionType{algorithm}

\begin{document}

\title{Phase Transitions in Neural Networks Pruning}

\author{Diego Pesce}
 \email{diego.pesce18@gmail.com}
\affiliation{%
 Dipartimento di Fisica, Universit\`a di Milano - Bicocca, 20126 Milano, Italy
}%
\affiliation{%
 ASC27,  via Tirso 14, 00198 Rome, Italy
}

 \author{Yang-Hui He}
\affiliation{%
London Institute for Mathematical Sciences (LIMS), Royal Institution, London W1S 4BS, UK}
\affiliation{
Merton College, University of Oxford, OX14JD, UK
}
\email{hey@maths.ox.ac.uk}
\author{Guido Caldarelli}

\affiliation{
Istituto dei Sistemi Complessi, CNR-ISC
}%
\affiliation{
DSMN and ECLT Ca' Foscari University of Venice
}%
\affiliation{
London Institute for Mathematical Sciences (LIMS), Royal Institution, London W1S 4BS, UK}%
\email{guido.caldarelli@unive.it}

\begin{abstract}
Deep neural networks are strongly over-parameterized, often containing far more weights than required for their task. Although such redundancy can aid optimization, it leads to inefficient deployment and high computational cost, motivating model compression techniques. Among these, network pruning provides a clear and effective route to sparsity.
We study pruning from a statistical-physics perspective, interpreting performance degradation under weight removal as a phase transition. Focusing on magnitude-based pruning with fine-tuning, we show that deep networks undergo a sharp transition from a cooperative, functional phase to a disordered phase with collapsed performance. This transition is characterized by scaling laws consistent with second-order critical behavior, with connectivity as the control parameter. Our findings suggest universal pruning-induced criticality across architectures and datasets. Finally, we show that  there exists a large class of subnetworks sharing the same nodes' degrees with similar learning ability, thus linking model performance to its topological properties.
\end{abstract}


\maketitle

% \tableofcontents

%\section{\label{sec:intro}Introduction}
\ac{ML} algorithms \cite{jordan2015machine}, particularly \ac{ANNs} \cite{yegnanarayana2009artificial}, are central tools for complex adaptive tasks. \ac{ANNs} originated with Rosenblatt’s perceptron \cite{rosenblatt1958perceptron}, which introduced learning through adaptive weights but was limited to linearly separable problems. This limitation was overcome in the 1980s with Hopfield networks \cite{hopfield1982neural}, connecting neural computation to collective dynamics, and multilayer architectures trained via backpropagation, notably advanced by Hinton and collaborators \cite{hinton1986learning}.

Modern standard \ac{ANNs} consist of layered interconnected units that aggregate weighted inputs and apply nonlinear activations. Training adjusts synaptic weights—typically through gradient-based optimization—enabling the approximation of complex functions for tasks such as classification and regression. Despite their empirical success, the principles governing \ac{ANNs} performance remain incompletely understood, often leading to overparameterized and unnecessarily computationally costly models. Although the benefits of overparameterization on optimization have been discussed \cite{buhai2020empirical}, the inefficiency of this approach lead to the development of model compression techniques \cite{neill2020overview,marino2023deep}. Among these, network pruning holds an leading role due to the simplicity in its implementation and its effectiveness in reducing the model's parameters without compromising its performance \cite{zhu2025comprehensive,cheng2024survey}.

We introduce a physics-inspired framework grounded in percolation theory \cite{stauffer2018introduction,bollobas2006percolation}, phase-transition analysis \cite{stanley1999scaling}, and network theory \cite{caldarelli2007scale} to quantify the interplay between architecture and algorithmic efficacy. By progressively pruning weak connections and retraining the remaining structure, we show that up to 98\% of edges can be removed with minimal performance loss.

This approach differs from the Lottery Ticket Hypothesis \cite{frankle2019lottery,malach2020proving,da2022proving}, which identifies sparse subnetworks—typically retaining about 10\% of the original parameters—that perform competitively when trained in isolation. Here, weights are continuously reorganized during pruning, independently of initialization, yielding functional subnetworks as small as 0.1\% of the original parameter count.
This self-organized process bridges statistical physics and deep learning, supporting the view of \ac{ML} models as interacting many-body systems evolving under an energy-like optimization dynamics \cite{anderson1972more}, and motivating the search for universal behavior analogous to percolation phenomena.
We validate the framework on two representative architectures: a four-layer fully connected network (\ac{FC}) and a five-layer convolutional network (\ac{CNN}). Both are trained on the MNIST and KMNIST datasets \cite{deng2012mnist,clanuwat2018deep}, which comprise ten balanced classes of handwritten symbols. After convergence under standard training protocols—cross-entropy loss, Adam optimization \cite{kingma2017adam}, and early stopping—the networks undergo iterative pruning and retraining until structural collapse. Tracking performance and structural observables reveals clear signatures of a pruning-induced phase transition.
To characterize the impact of pruning on network functionality and confidence, we monitor the following observables.

{\em \ac{ACC}} is defined as the fraction of correct predictions and serves as a proxy for functional performance. As pruning progresses, accuracy typically remains nearly constant in the initial stages and then drops sharply once structurally relevant connections are removed. This behavior provides a direct measure of network robustness and signals the onset of critical degradation. Baseline accuracies in MNIST are 97.03\% and  99.05\%  for \ac{FC} and \ac{CNN} respectively; the same values for KMNIST are 86.06\% and 92.52\%. \ac{CNN} outperforms constantly \ac{FC}, particularly on the more complex KMNIST dataset.

{\em \ac{PE}}, defined below, quantifies diversity of predicted classes, independently of correctness:
\begin{equation}
PE = -\sum_{i=1}^{K} f_i \log f_i ,
\end{equation}
where $f_i$ denotes the fraction of test samples assigned to class $i$. For a balanced dataset, a properly functioning network yields high prediction entropy, corresponding to an approximately uniform use of all classes. Marked deviations or abrupt reductions signal degraded feature extraction, class collapse, or structural disconnection.

\begin{figure}
    \centering
    \subfloat[\ac{FC} architecture on MNIST dataset.]{
    \includegraphics[width=0.95\linewidth]{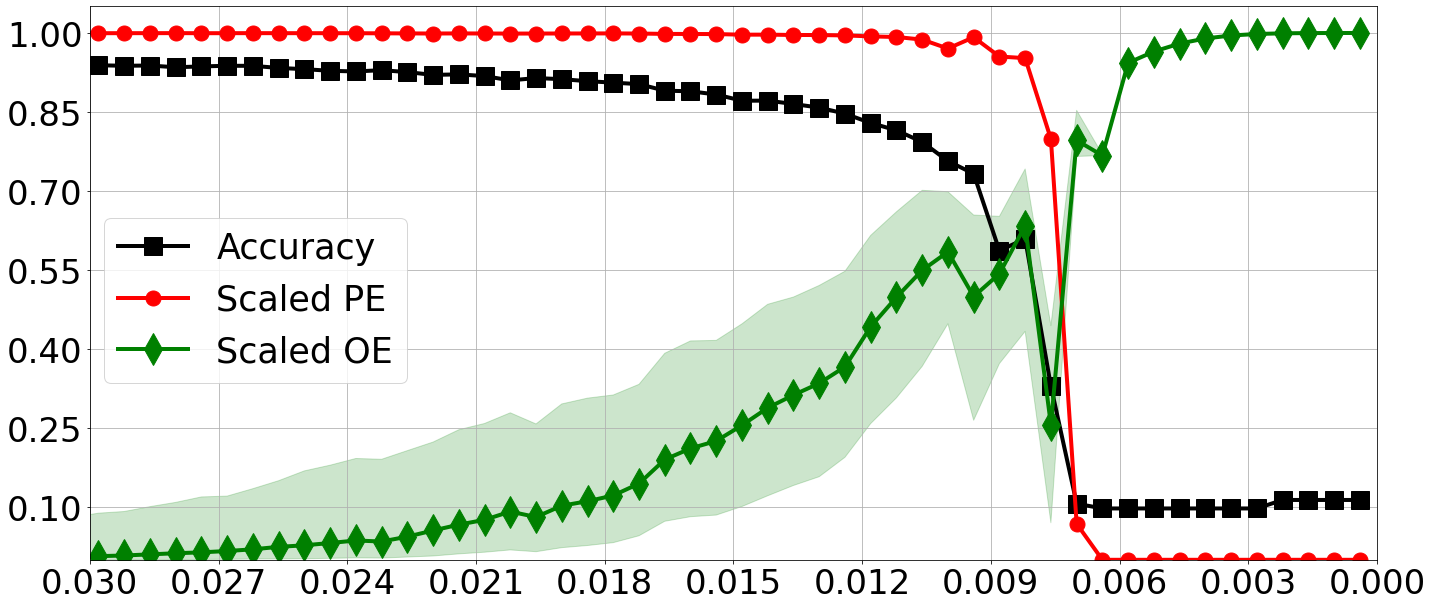}
    }

    \subfloat[\ac{FC} architecture on KMNIST dataset.]{
    \includegraphics[width=0.95\linewidth]{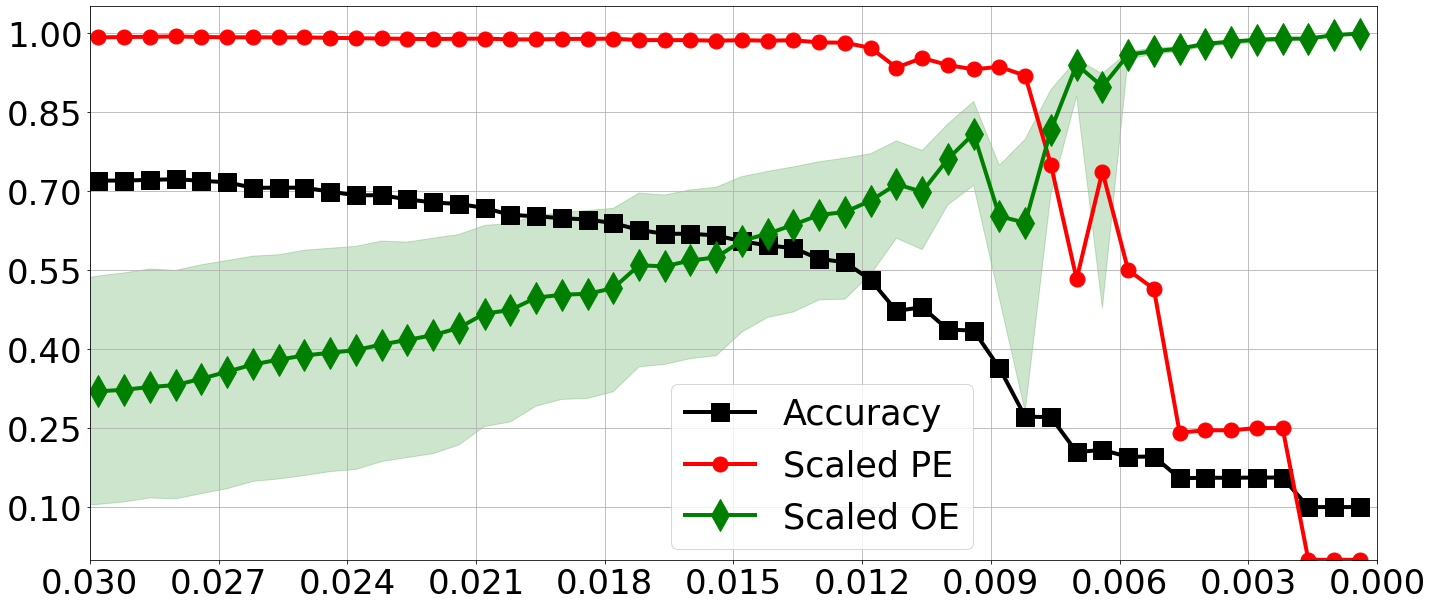}
    }

\caption{(color on line) Detail of the behavior near the point at which the algorithm ceases to operate. The x axis shows the percentage of links remaining after pruning. From top to bottom: accuracy (black line), prediction entropy (red line) and outcome entropy (green line). Results are shown for the \ac{FC} architecture using (a) MNIST and (b) KMNIST.}
\label{f:NN_smallest_zoom}
\end{figure}

{\em \ac{OE}} measures the confidence of the network’s predictions and is distinct from prediction entropy. A sharply peaked softmax output corresponds to low entropy and high confidence, whereas a nearly uniform output indicates uncertainty. Averaged over the test set 
$E$, outcome entropy is defined as
\begin{equation}
OE = \mathrm{Ave}_{x \in E}\left[-\sum_i p_i(x)\log p_i(x)\right] ,
\end{equation}
where 
$p_i(x)$ is the softmax probability assigned to class 
$i$ for input $x$. This observable typically increases as pruning degrades the network’s ability to discriminate between classes. Note that both $\ac{PE}$ and $\ac{OE}$ range from $0$ to $\log(10)$ in this case.
Figure~\ref{f:NN_smallest_zoom} shows accuracy, prediction entropy, and outcome entropy versus pruning level, defined as the fraction of remaining connections. Pruning is performed in 5,000 equally spaced steps, providing high-resolution dynamics while remaining computationally feasible (see Supplementary Materials for details).

For \ac{FC}, two regimes are observed. In Fig.~1(a), a percolation-like transition occurs below 1\% of remaining edges. In Fig.~1(b), the degradation is smoother, with functionality sustained until a common threshold and a sharp collapse near 1.2\% of the original edges. In both cases, KMNIST deteriorates earlier and more gradually than MNIST, reflecting its greater complexity. \ac{CNN} exhibits qualitatively similar behavior (see Supplemental Materials), with KMNIST again showing earlier and smoother degradation.

%\subsection{\label{subsec:crit_exp} Scaling laws}
Visual inspection already suggests a sharp transition between functioning and nonfunctioning regimes. We therefore analyze this change within the framework of percolation phase transitions, studying accuracy and prediction entropy as functions of the distance of the edge density $p$ from the critical value $p_c$
\begin{eqnarray}
    A &\propto & (p-p_c)^{\mu} \\
    PE& \propto & (p-p_c)^{\theta}
\end{eqnarray}
$\mu \text{ and } \theta$ are the critical exponents.
In the Supplementary Materials we report a table of the estimated critical exponents, noting that in the \ac{FC}–KMNIST case statistical fluctuations prevent a reliable determination. Despite finite-size effects and architecture-specific features, the exponents are qualitatively consistent across order parameters, suggesting a common universality class under the present pruning protocol.

Tracking weight evolution reveals smooth dynamics throughout most of the process, consistent with the preservation of stable configurations as small weights are progressively removed. Pronounced peaks in the \ac{EM} arise only near the operational limit (Fig.~\ref{fig:NN_MNIST_phases}), indicating that structured pruning isolates the network’s essential substructure more effectively than random removal, at least in the systems analyzed. The strongest rearrangements occur in the final layer, whose parameters adapt to compensate for changes in the internal data representation.

    \begin{figure}
    \includegraphics[width=0.33\textwidth]{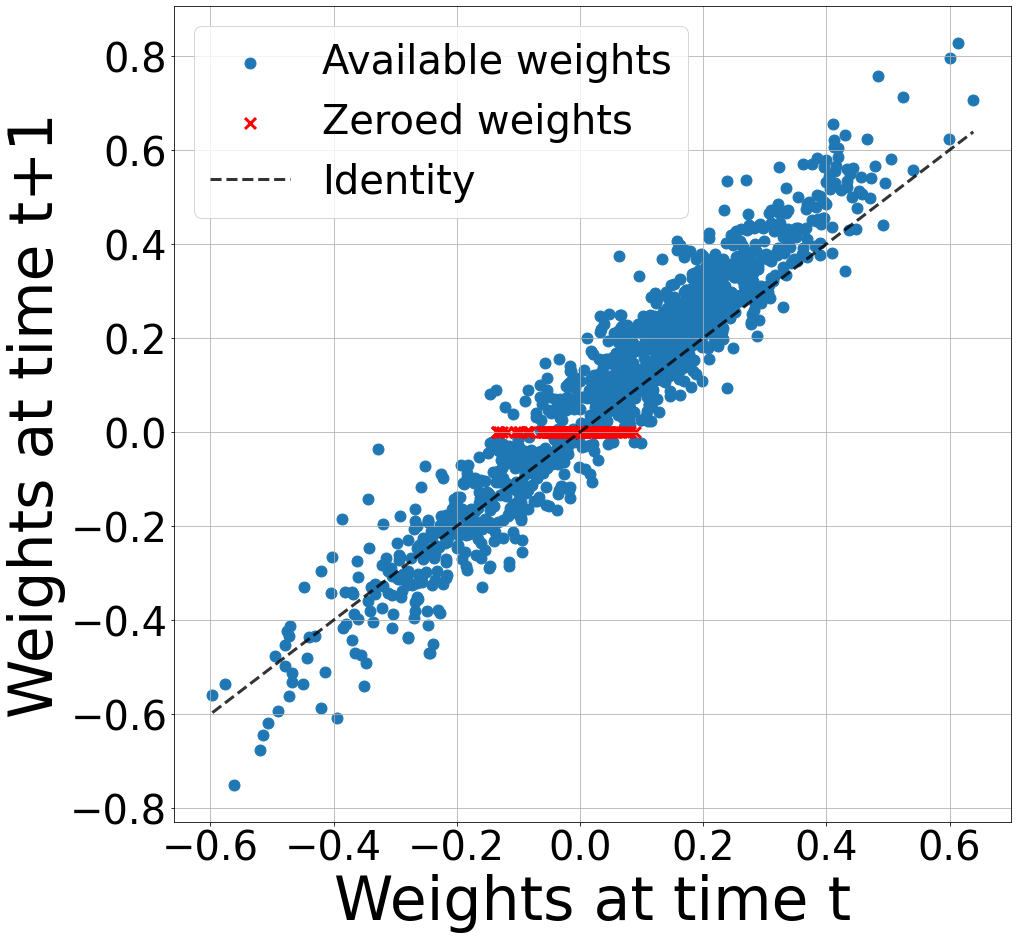}
    \caption{\label{fig:weights_evol} Evolution of the weights over multiple magnitude-based pruning steps. The y axis shows the values of the remaining parameters, plotted against their values at the previous timestep (x axis). Red points denote pruned parameters, while blue points show surviving weights. The deviation from the identity line indicates weight evolution between steps.}
    \end{figure}
    Figure \ref{fig:weights_evol} shows weights evolution over a jump of 100 pruning steps. Red crosses indicate parameters pruned during the jump. Excluding these points, the remaining weights closely follow the identity line, implying negligible evolution. Deviations from identity, measured by the $R^2$ coefficient, quantify weights updates. 
    $$
    R^2(w_t, w_{t'}) = 1-\frac{RSS}{TSS} = 1-\frac{\lVert w_{t'}-w_t\rVert^2}{\lVert w_{t'}-\bar w_{t'}\rVert^2}
    $$
    
where ($\bar w$) denotes the component-wise average of (w), and RSS and TSS are the residual and total sum of squares, respectively. As pruning progresses, the number of nonzero weights (blue points in Fig. \ref{fig:weights_evol}) decreases, making $R^2$ values across timesteps not directly comparable. We therefore use the adjusted-$R^2$, which accounts for the changing number of data points.
$$
\text{Adj-}R^2(w_t, w_{t'} | n) = 1- \left(\frac{n-1}{n-2}\right)\frac{\lVert w_{t'}-w_t\rVert}{\lVert w_{t'}-\bar w_{t'}\rVert}^2
$$
where $n$ is the amount of the remaining (blue) points after the deletion of the parameters.
Note that this coefficient is not symmetric: prediction and target must be specified and are not interchangeable. Here $w_{t'}$ is the target. Since we are interested in how weights at a given timestep differ from those at an earlier one, a prediction–target interpretation naturally arises.
Finally, \ac{EM} reads as follows
\begin{align*}
\text{EM}(w_t, w_t' | n) &= \Theta(t-t')\left[1-\text{Adj-}R^2(w_t, w_{t'} | n)\right] \\ &+ \Theta(t'-t)\left[1-\text{Adj-}R^2(w_{t'}, w_t | n)\right] 
\end{align*}   
where $\Theta$ is the standard Heaviside's function.
% $$
% \text{EM}(w_t, w_t' | n) = 
%     \begin{cases}
%         1-\text{Adj-}R^2(w_t, w_{t'} | n) \\ \hspace{20pt} = \left(\frac{n-1}{n-2}\right)\left(\frac{\lVert w_{t'}-w_t\rVert}{\lVert w_{t'}-\bar w_{t'}\rVert}\right)^2 & \text{if } \hspace{2pt} t' < t \\
        
%         \\
        
%         1-\text{Adj-}R^2(w_{t'}, w_t | n) \\ \hspace{20pt} =\left(\frac{n-1}{n-2}\right)\left(\frac{\lVert w_{t'}-w_t\rVert}{\lVert w_t-\bar w_t\rVert}\right)^2 & \text{if } \hspace{2pt} t' \geq t
%     \end{cases}
% $$
This \ac{EM} will allow us to quantitatively study the evolution of the weights. In turn, it will help in identifying critical regions of strong reorganizations separating two different behavioural regimes of the algorithms.

Fine-tuning after each pruning step is essential, as it enables the network to redistribute learning among the remaining parameters and partially compensate for removed connections. Without fine-tuning, performance degrades rapidly and critical behavior is obscured. Training is performed with a pruning mask, keeping removed weights fixed and their gradients zeroed.

This procedure reveals a computational backbone: a sparse subnetwork that preserves predictive performance. While the existence of such subnetworks was previously noted \cite{frankle2019lottery}, magnitude-based pruning with fine-tuning exposes this backbone in a controlled manner, enabling a detailed analysis of the network’s phase behavior.
\begin{figure}
    \includegraphics[width=0.5\textwidth]{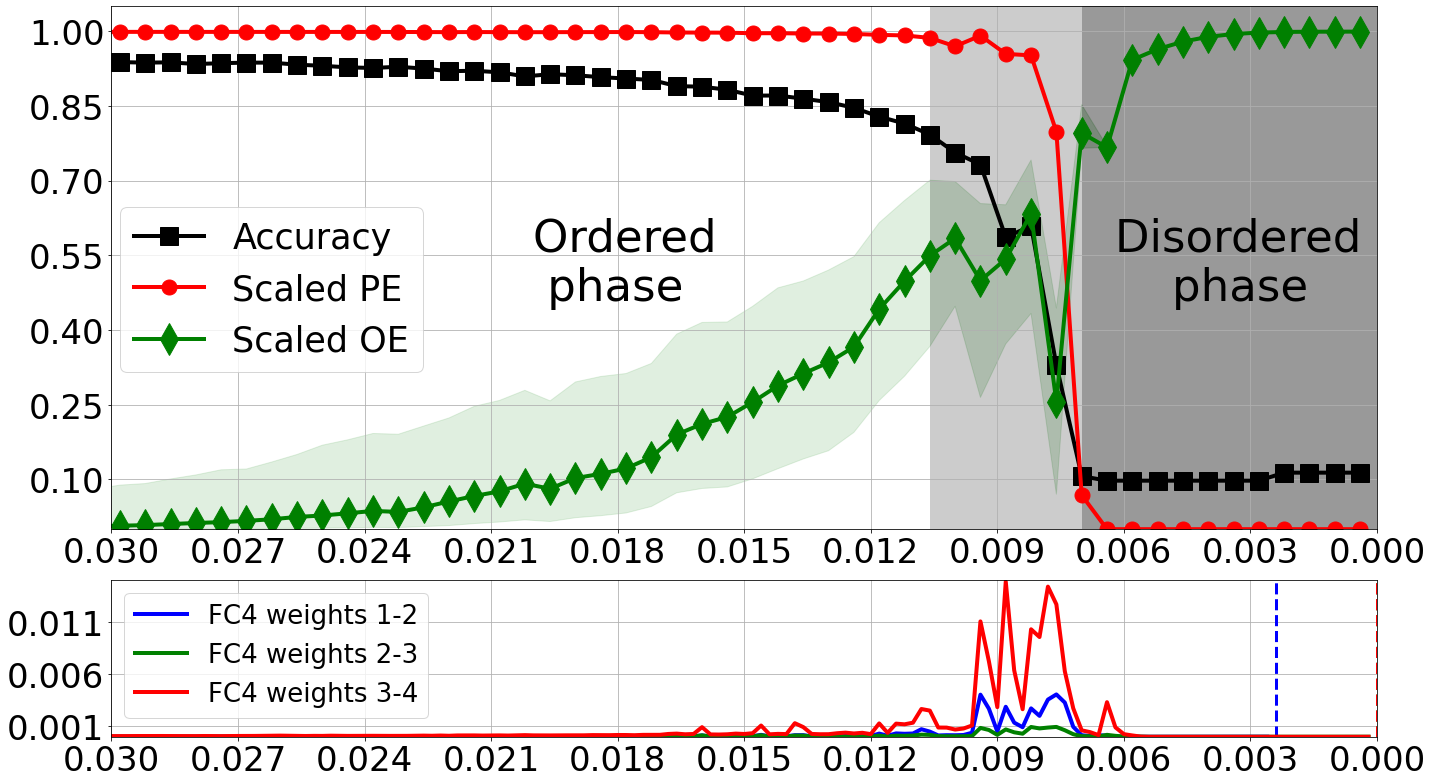}
    \caption{\label{fig:NN_MNIST_phases} The plot shows the various order parameters compared and the corresponding evolution metrics for weights. It is evident that the region of stronger reorganization is the one signing the lost of model performances.}
\end{figure}

Analogously to spins in the Ising model, \ac{ANNs} undergo a transition from an ordered, cooperative phase—where a sufficient fraction of weights supports reliable discrimination—to a disordered regime in which pruning progressively destroys coherence. As weights reorganize, long-range correlations responsible for task-specific structure vanish, driving the network toward a more stable but task-agnostic configuration (Fig. \ref{fig:NN_MNIST_phases}).
This transition can be interpreted as symmetry restoration: in the disordered phase, class-conditional outputs average to a uniform distribution, reflecting the loss of class-specific organization. Although individual predictions may remain nonuniform, their class average becomes uniform, indicating preserved expressivity but no meaningful task representation. Notably, even after accuracy and prediction entropy plateau at minimal values, the outcome entropy remains unsaturated, suggesting that residual input–output information pathways persist and that gradient-based recovery is, in principle, still possible.

Finally, we investigate the rationale behind the LLM’s functioning after pruning. As {\em starting configurations}, we consider randomizations of the experimentally obtained structure, generated by selecting pairs of edges and exchanging (when possible) their end vertices. Training is then performed on these structures, initializing weights by pruning the full-network initial weights accordingly. Results are compared with equally sized randomly pruned networks subject to the same initialization; in this case, however, network properties—particularly nodes’ degree—do not arise from the self-organized pruning procedure.

\begin{figure}
\includegraphics[width=0.36\textwidth]{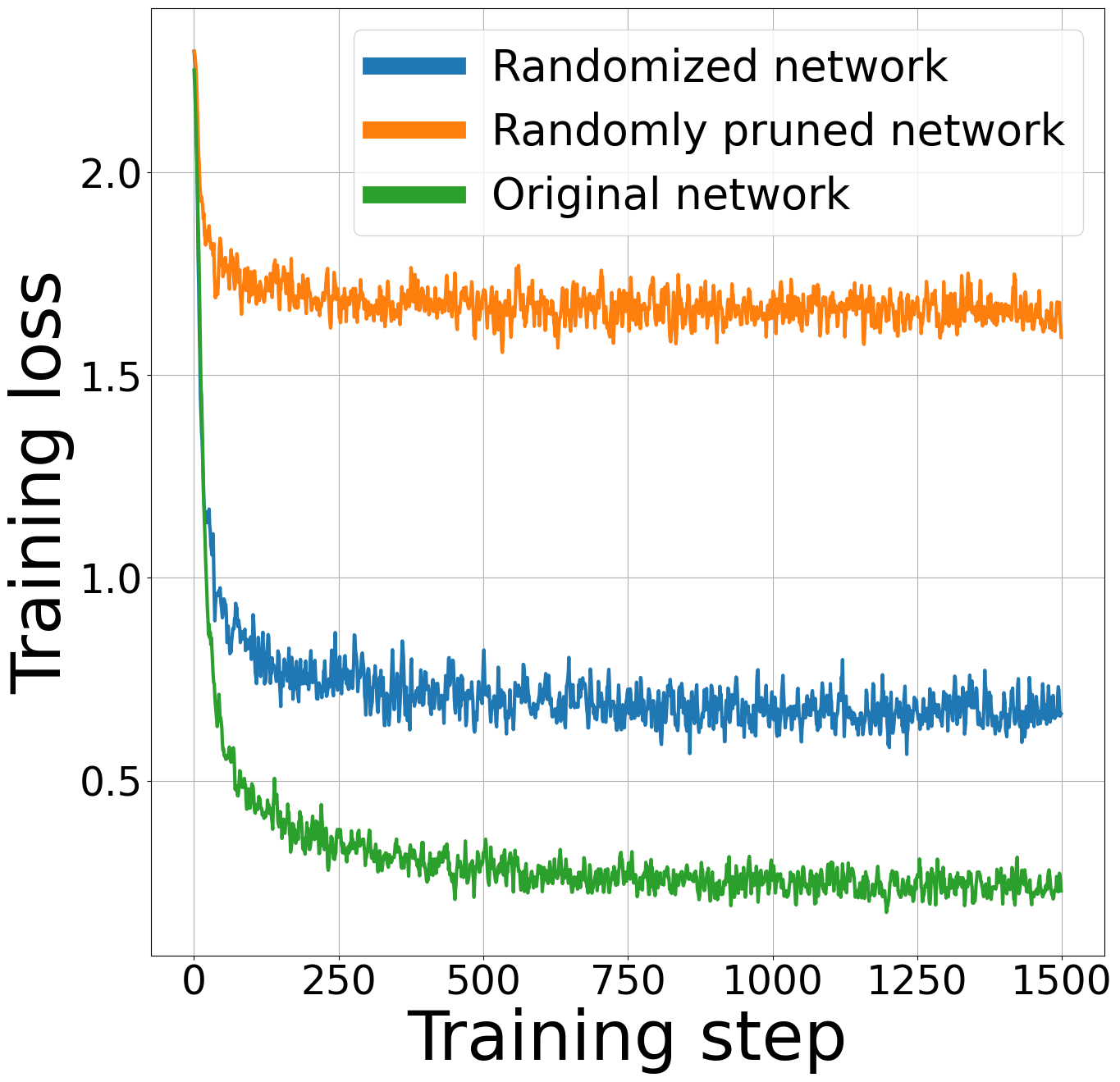}
\caption{\label{fig:comparison}(color on line) Here we show the loss in the training on y-axis with respect to the steps of training on x axis. From top to down we have (a; orange online) totally random subset of edges, (b; blue online) subset of edges self-organised in the pruning procedure and then randomized, (c; green online) the original model }
 \end{figure}

As shown in Fig.\ref{fig:comparison}, although pruned subnetworks achieve lower efficacy than the original model, the fully random ones cannot be trained. These findings link learning capability to topological properties, especially nodes’ degree.

In conclusion, pruning in simple \ac{ANNs} architectures can be interpreted as a phase-transition phenomenon with universal features largely independent of architecture and dataset. This framework unifies the characterization of robustness and functional collapse, with implications for more complex models. Our procedure highlights a self-organization of weights that preserves acceptable performance despite severe parameter reduction, enabling immediate and substantial computational savings and fostering more efficient, sustainable machine learning implementations. Further analysis of pruned subnetworks topology may guide the design of efficient, application-tailored networks and ease the training procedures.

\bibliography{bibliography}% Produces the bibliography via BibTeX.

\end{document}